\documentclass[twocolumn,showpacs,preprintnumbers,amsmath,amssymb]{revtex4}
  \usepackage{graphicx}
  \usepackage{dcolumn}
  \usepackage{bm}
 
\begin{document}

\title{The behavior of f-levels in HCP and BCC rare-earth elements
in the ground state and in XPS and BIS spectroscopy from density-functional theory.}

  \author{T. Jarlborg}

  \affiliation{
  DPMC, University of Geneva, 24 Quai Ernest-Ansermet, CH-1211 Geneva 4,
  Switzerland
  \\
  }
 
 
\begin{abstract}
The electronic structures of rare-earth elements in the HCP structure, and Europium in the BCC
structure, are calculated by use of density-functional theory, DFT. Simulation of X-ray
photoemission spectroscopy (XPS) and bremsstrahlung isochromatic spectroscopy
(BIS) are made within DFT by imposing that f-electrons are 
excited by a large photon energy, either by removing from the occupied states
in XPS, or by adding to the unoccupied f-states in BIS. The results show sizable 
differences in the apparent position of the f-states compared to the f-band energy
of the ground states. This result is fundamentally different from calculations 
assuming strong on-site correlation since all calculations are based on
DFT.
Spin-orbit coupling and multiplet splittings are not included. The 
present simulation accounts for almost half of the difference between
the f-level positions in the DFT ground states and the observed
f-level positions. The electronic specific heat at low $T$
is compatible with the DFT ground state, where f-electrons often reside at the Fermi level.
  \end{abstract}
 
  \pacs{71.20.Eh,
        71.28.+d,
        79.60.-i}
 
  \maketitle

\section{Introduction}

Partially filled f-orbitals are 
 predicted by the density-functional theory (DFT) for the ground state
to be contained in narrow bands 
with a high density-of-states (DOS) at the Fermi energy $E_F$.  
The fact that the $\alpha-\gamma$-transition in fcc Ce can be described quite accurately
by temperature dependent DFT calculations in which vibrational, 
electronic and magnetic free energies are
taken into account \cite{ce}, shows that DFT is more 
reliable than what can be expected for f-electron
systems.
However, spectroscopic signatures of f bands are often found 
several eV's above 
or below the $E_F$ depending on the nature of the spectroscopy \cite{lang},
and the main weight is not the Fermi energy $E_F$, 
as DFT predicts for the ground state in most of the rare-earths. 
Atomic calculations with imposed occupations of the 4f-orbitals \cite{herb,joha}, 
based on the assumption of strong electronic correlations among localized electrons,
have been used for interpretation of spectroscopic data
\cite{lang,marel}. Atomic levels are split and disconnected from $E_F$ by the on-site
correlation, represented by a Hubbard parameter $U$, 
but fundamental questions arise about what happens when the f-electrons form bands in metallic
solids, and about the real nature of the ground state.
These problems may be elucidated by a DFT approach 
\cite{ncco}, which is tailored for the precise spectroscopic probe by including relaxation energies 
relevant for excitations between occupied and empty bands.
The calculations are in the spirit of
the final state rule (FSR), which basically assumes that a system can relax around its final state
configuration before the emission/absorption of a photon \cite{fsr}.

In the present work we apply the relaxation  
approach \cite{ncco} to several of the 4f-electron rare-earth (RE)
lanthanides in order to search for a physically acceptable description of X-ray 
photoemission spectroscopy (XPS)
and bremsstrahlung isochromat spectroscopy (BIS).
We are not seeking for an agreement with measured intensities,
since inclusion of matrix
elements, multiplets and spin-orbit (SO) interaction would be needed for that. But we determine the energy
renormalization of the f-bands in order to see if they can lead to a better reconciliation
between the DFT ground state band positions and the center of gravity of the spectroscopic f-band peaks.
The goal is to apply a similar method as the method used for excitations of core electrons, where the threshold
energies in X-ray absorption spectroscopy (XAS) are much improved over unrelaxed core level energies in metal silicates  
\cite{lerch}. The excited core electron is in those XAS calculations added to the valence electrons,
leaving a core hole behind. However, core electrons are localized and atomic-like methods can be
applied. Here, in the present approach for XPS
and BIS further considerations are needed for
transitions between delocalized and hybridized valence states 
and the continuum at energies of the order $\sim \hbar \omega$  above $E_F$.

From the results of this work it is suggested that DFT is essentially correct for 4f-levels in RE elements, but
that spectroscopic data have been interpreted incorrectly about the signatures of the f-band centers
far from $E_F$. Correlation in the ground state
 is not the source of seeing f-levels far form $E_F$, but
screening in the excitation process makes it look that way. This conclusion is corroborated by electronic
specific heat calculations, which are compatible with experiments if the f-DOS is large at $E_F$.

An outline of this paper is as follows. In Sec. II, we present 
the  details of the DFT computations and total energies for ground state and excited state configurations. 
The results of the calculations are presented and 
compared with experimental results in Sec. III, together with the results of electronic specific
heat calculations. The conclusions are given in Sec. IV.

\section{Method of calculation}

Self-consistent density-functional Linear Muffin-Tin Orbital (LMTO) band calculations \cite{lmto}
are made for the ground state and exited states
in hcp and bcc rare-earth elements using  potentials based on
the local spin-density approximation, LSDA \cite{lsda}. The excited states involve
the localized 4f states, and in order to avoid interaction between
excitations on neighboring atoms we consider 16 atom supercells where only one atom is excited. 
The 16 atom supercells are made by
doubling the ordinary hcp unit cell and the cubic bcc double cell in each direction
($x,y,z$), respectively.
The self-consistent convergence is obtained
using a mesh of 30 or 35 k-points
within the irreducible Brillouin zones corresponding to hcp or
bcc 16-atom supercells, respectively. The c/a ratio is taken to be the
same for all hcp structures, 1.59, which is a fair average for the
different systems.
The lattice constants $a_0$ for each system are close to
the experimental ones given in ref. \cite{kitt}.
The complex structure of Sm and the fcc structure Yb
are approximated by the hcp structure in these calculations. It is not expected that
the excitation energies depend strongly on the c/a-ratios or the exact
structures. However, as will be discussed, the excitations depend much on the f-band occupation
and spin polarization.
All calculations are spin-polarized. 
The deep 5p states are always included as band states. Together with the 4f-states they
are very localized with narrow band widths. The LMTO linearization energies are chosen within
the band region, i.e. with negative logarithmic derivatives. Self-consistency is more delicate than usual
because of the removal/addition of electrons within the narrow bands. The remaining f-bands
on the excited atom are sometimes moving in energy during the iterations, and 
the linearization energy is then adjusted to follow the f-band center.
The method of calculation is close to what has been used earlier for the Nd-f band in electron
doped Nd$_2$CuO$_4$ \cite{ncco}. 
Calculations are made where a fraction ($\delta$) of an electron is excited
in order to focus on excitations from precise peaks in the DOS. A high precision of the
calculated total energies, $E_T^{\delta}$, is needed, and
the relaxation energies, $\Delta \epsilon$, are defined per excited electron as $(E_T^0-E_T^{\delta})/\delta$.
Non-linearities
can appear if band edges interfere near $E_F$ for large $\delta$. 
Because of small technical differences between the computation codes for excited and ground state configurations,
it is more precise to calculate the
the total energies for the ground state, $E_T^0$, as $E_T^{\delta}$ for $\delta \rightarrow 0$ directly from the
excited state code.

Relaxation for excitations involving
localized f-electrons is expected to be more important than for 
excitations of itinerant electrons. The reason is the different radial shapes between f-states
and itinerant states. High-energy final states are itinerant, so transitions
to/from a localized f-state implies important reshaping of the charges, while this is not so for
transitions to/from a state which already is delocalized. Thus,
the charge is (in XPS) removed at an energy within the occupied 4f
majority state on one of the atoms and is spread out uniformly 
over the cell to account for a final
state at high energy \cite{note1}. The difference in total energy between this state and the
ground state defines the relaxation energy, $\Delta \epsilon$. Thus the final state XPS image will 
appear to have its f-peak shifted  
by an amount $\Delta \epsilon$ with respect to the Fermi level.
The excitation energy according to
the Koopmans approximation \cite{koop} would be the difference in ground state energy levels.
For instance in XPS $\hbar \omega$ would be equal to $E^f-E^i$, where $E^f$ and $E^i$ are the ground state
band energies for the final and initial states, respectively. The latter are the calculated LMTO eigenvalues,
but the energies $E^f$ are too large to be calculated by the LMTO code. However, the density of energy levels
at such high energy is large enough so that a final state energy level can always be found, independent
of k-point conservation \cite{jn}. Therefore, the Koopmans approximation would simply mean that 
the XPS spectrum would look like the occupied ground state DOS (all states shifted equally by $\hbar\omega$).   
 The renormalized excitation energy is corrected by the relaxation energy, $\hbar \omega - \Delta \epsilon$.
 Thus, according to the FSR there is time for the system to relax around the missing f-electron 
 before the excited electron can enter
 a level at high energy. The photon energy $\hbar\omega$, corrected by the relaxation energy
 $\Delta\epsilon$, will be given to the electron.

 In the XPS excited state simulation, the 
fractional electron charge $\delta$ is removed from
the local DOS on one site $t'$. This charge defines
an energy interval [$E_b,E_a$] on the local DOS; 
\begin{equation}
\sum_{\ell}\int_{E_b}^{E_a}N_{t',\ell}(E)=\delta
\end{equation}
where $N_{t',\ell}(E)$ is the DOS on the site $t'$ of character $\ell$.

 This charge is distributed uniformly within
the entire unit cell, of volume $\Omega$, in form of a charge density $\delta/\Omega$.
  The justification for this is that
for a very high excitation energy as for XPS, $E_f \approx \hbar\omega$, it is possible to ignore
the crystal potential $V(r)$ in comparison to $E_f$ \cite{jn}, when the Schr\"{o}dinger equation

\begin{equation}
(-{\nabla}^2 + V(r))\Psi(E_f,r) = E_f \Psi(E_f,r)
\label{sch1}
\end{equation}
is simplified to
\begin{equation}
  -{\nabla}^2 J(E_f,r) = E_f J(E_f,r)
\label{sch2}
\end{equation}
where the free-electron solutions $J(E_f,r) \sim exp(i \sqrt{(E_f) \cdot r)}$ have a constant density.

The total charge density for the excited state is then;
\begin{eqnarray}
\rho(r)=\sum_{t,\ell}\int_{-\infty}^{E_F}N_{t,\ell}R^2_{t,\ell}(E,r)dE \nonumber \\
-  \sum_{\ell}\int_{E_b}^{E_a}N_{t',\ell}(E)R^2_{t',\ell}(E,r)dE +\delta/\Omega
\label{rho}
\end{eqnarray}
where $R_{t',\ell}(E,r)$ are the radial wave functions and $E_a \leq E_F$.
The total energy $E_T = U_C + T + E_{xc}$, where the Coulomb energy $U_C$ and exchange-correlation
energy $E_{xc}$ are calculated using the constrained density from eq. \ref{rho} and kinetic energy $T$ is:
\begin{equation}
T(\delta)=\sum_{t,\ell}\int_{-\infty}^{E_F} E N_{t,\ell}(E)dE \nonumber \\
-  \sum_{\ell}\int_{E_b}^{E_a} E N_{t',\ell}(E)dE + \hbar\omega\delta
\label{ekin}
\end{equation}

The self-consistent field (SCF) iterations are repeated while keeping the [$E_b,E_a$] interval at the
4f band until the
total energy is converged.
The energy interval [$E_b,E_a$] is narrow in all rare-earth elements
because of their high 4f-band DOS. The removed charge is mostly of pure
f-character, because the f-DOS is much larger than other $\ell$-DOS (exceptions are Yb and Lu, where 
5p is in the same energy range as 4f).
The SCF procedure with excitation is less stable than for ordinary ground state calculations, and
it is often difficult or slow to achieve convergence.

The method for the inverse procedure, for BIS, is modified so that the fraction of an electronic charge
is added within an energy interval in the empty 4f band above $E_F$, and the compensating charge
density $\delta/\Omega$ is removed everywhere. No BIS calculations were made for Yb and Lu, since they have
no empty f-states. All other elements have empty f-states in the minority bands for which calculations 
are made. 
For XPS only calculations for excitations from the majority spins were considered here.

\section{Results}

\subsection{Ground state}

\begin{figure}[h]
  \begin{center}
  \includegraphics[width=8cm,height=6cm]{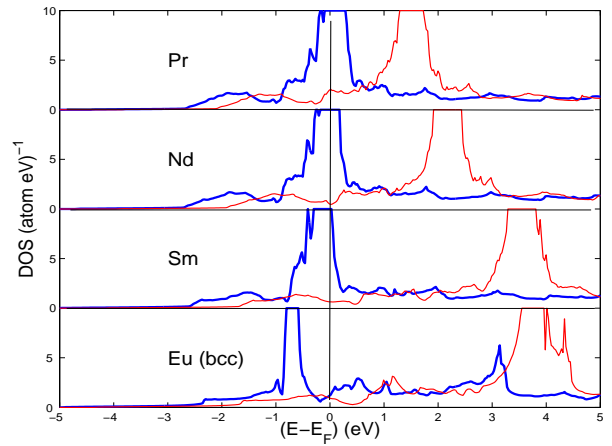}
  \end{center}
  \caption{Ground state majority, bold (blue) line, and minority, thin (red) line, DOS functions
  for Pr, Nd, Sm and Eu. All functions are cut at $N(E)$=10 states/atom eV.} 
  \label{figdos1}
  \end{figure}

\begin{figure}[h]
  \begin{center}
  \includegraphics[width=8cm,height=6cm]{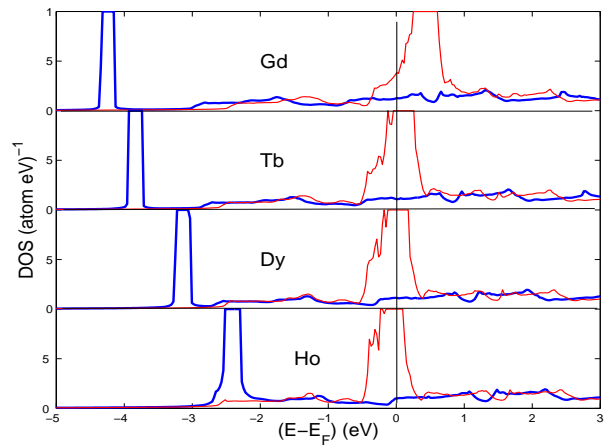}
  \end{center}
  \caption{Ground state DOS for Gd, Tb, Dy and Ho, presented as in Fig. \ref{figdos1}.} 
  \label{figdos2}
  \end{figure}

\begin{figure}[h]
  \begin{center}
  \includegraphics[width=8cm,height=6cm]{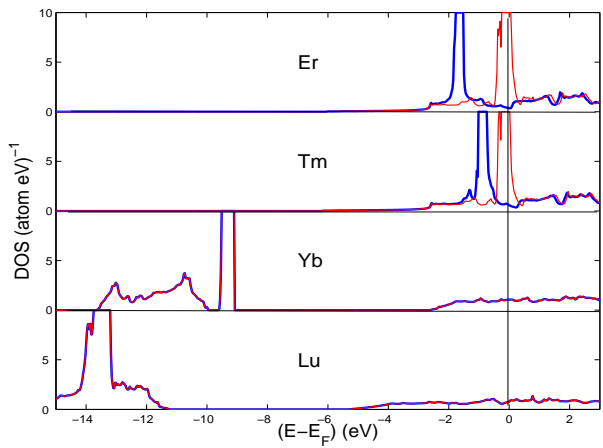}
  \end{center}
  \caption{Ground state DOS for Er, Tm, Yb and Lu, presented as in Fig. \ref{figdos1}.
  The wide parts of the DOS near the 4f peaks in Yb and Lu are due to 5p states.} 
  \label{figdos3}
  \end{figure}

Results from the ground state calculations are summarized in Table I and Figures \ref{figdos1}-\ref{figdos3}.
Several band calculations are found in the literature for Gd \cite{kub,tem,sing,byl}, where
the accuracy of LSDA potentials and the sensitivity to basis functions are discussed.
The present result agree well with the other calculations without SO-coupling concerning the f-band positions
and the band widths. It is difficult to obtain good values for the f-band energies, and 
the f-levels are often treated separately from the valence electrons, as in calculations for Sr and Yb \cite{kubo}.

The high $N_{\uparrow}(E_F)$-values for Pr, Nd and Sm show that $E_F$ crosses the 
f-band. For Tb, Dy, Ho, Er and Tm the Fermi levels cross the minority band,
as can be concluded from their high $N_{\downarrow}(E_F)$-values. The bottom of the minority band
in Gd is very near $E_F$, and its $N_{\downarrow}(E_F)$-value is only moderately large.
The partial occupations of majority/minority bands explain the variation of spin moments
among the different RE elements. The magnetic moments follow closely Hund's first rule.
For instance, all f-electrons are polarized in Eu and Gd, and $Q_f \approx m$.
The majority and minority bands in Yb and Lu are degenerate and completely filled, 
with no exchange splitting and no moment.

\begin{table}[ht]
\caption{\label{table1}
Summary of the LDA ground state results.
Rare-earth element and structure, lattice constant, number of valence electrons ($Q$), number of f-electrons ($Q_f$) and
magnetic moment ($m$) per site, calculated differences between $E_F$ and the center of gravity
of the occupied and unoccupied f-DOS, ($\epsilon_{occ}$ and $\epsilon_{un}$, respectively), and DOS at $E_F$ for
majority and minority spin, respectively (in states/atom/eV).}
  \vskip 2mm
  \begin{center}
  \begin{tabular}{l c c c c c c c c}
  \hline

 RE (str.) & $a_0 (\AA)$ & $Q$ & $Q_f$ & $m$  & $\epsilon_{occ}$ & $\epsilon_{un}$ & $N_{\uparrow}$ &  $N_{\downarrow}$\\

  \hline \hline

Pr (hcp) & 3.67 & 11 & 2.49 & 2.66 & -0.2 & 1.5 & 10.9 & 1.0 \\
Nd (hcp) & 3.66 & 12 & 3.64 & 4.03 & -0.3 & 2.2 & 15.4 & 0.24 \\
Sm (hcp) & 3.67 & 14 & 5.86 & 6.26 & -0.4 & 3.5 & 14.1 & 0.32 \\
Eu (bcc) & 4.58 & 15 & 6.94 & 7.23 & -0.9 & 3.8 & 0.67 & 0.45 \\
Gd (hcp) & 3.63 & 16 & 7.31 & 6.95 & -4.2 & 0.4 & 0.63 & 1.92 \\
Tb (hcp) & 3.60 & 17 & 8.45 & 5.58 & -3.8 & 0.2 & 0.55 & 6.54 \\
Dy (hcp) & 3.59 & 18 & 9.59 & 4.40 & -3.2 & 0.2 & 0.65 & 15.3 \\
Ho (hcp) & 3.58 & 19 & 10.72 & 3.19 & -2.4 & 0.1 & 0.48 & 28.0 \\
Er (hcp) & 3.56 & 20 & 11.82 & 2.04 & -1.7 & 0.1 & 0.22 & 34.2 \\
Tm (hcp) & 3.54 & 21 & 12.89 & 0.96 & -0.8 & 0.1 & 0.29 & 19.5 \\
Yb (hcp) & 3.91 & 22 & 14.00 & 0.0 & -9.3 & 0.0 & 0.53 & 0.53 \\
Lu (hcp) & 3.50 & 23 & 14.00 & 0.0 & -13.5 & 0.0 & 0.41 & 0.41 \\

  \hline
  \end{tabular}
  \end{center}
  \end{table}

\subsection{Energy renormalization}

\begin{figure}[h]
  \begin{center}
  \includegraphics[width=9cm,height=7cm]{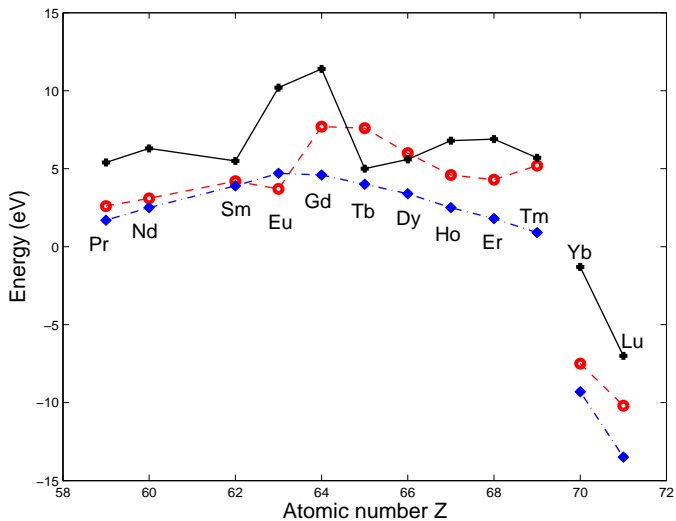}
  \end{center}
  \caption{Calculated energy difference between the unoccupied and occupied f-band centers (blue diamonds
  connected by a semi-broken line),
  calculated differences in BIS and XPS peak positions (red circles connected by a broken line) and what is called 
  "correlation energy $U$" (black plus-signs connected by a thin line)
  from the measurements of Lang {\it et al} \cite{lang}. For Yb and Lu only the unrenormalized and normalized
  occupied XPS values are shown and compared to the corresponding parameter "$\Delta_-$" from
  the experimental paper \cite{lang}. Note that the results indicated by the diamonds have no
  particular correlation beyond LSDA. As is explained in the text, the experimental values should not be assigned 
  to correlation $U$. 
  The results indicated by the circles include relaxation proper to the XPS and BIS processes,
  and it improves generally the comparison with experimental peak positions.} 
  \label{fig4}
  \end{figure}

The results of the XPS and BIS excitation energy per electron 
are given in Table \ref{table2}
and reproduce similar trends to those given in Ref.~\cite{ncco} 
for Nd$_{2-x}$Ce$_x$CuO$_4$.
The important reference for experimental comparison is the work by Lang {\it et al} \cite{lang},
which provides detailed information about the measured XPS and BIS intensities
in the RE-elements, as well as numbers for what they
believe are due to correlation. In Fig. \ref{fig4} is a summary of the combined XPS and BIS simulations
with comparison to experimental values of $U$ \cite{lang}.

\begin{figure}[h]
  \begin{center}
  \includegraphics[width=9cm,height=7cm]{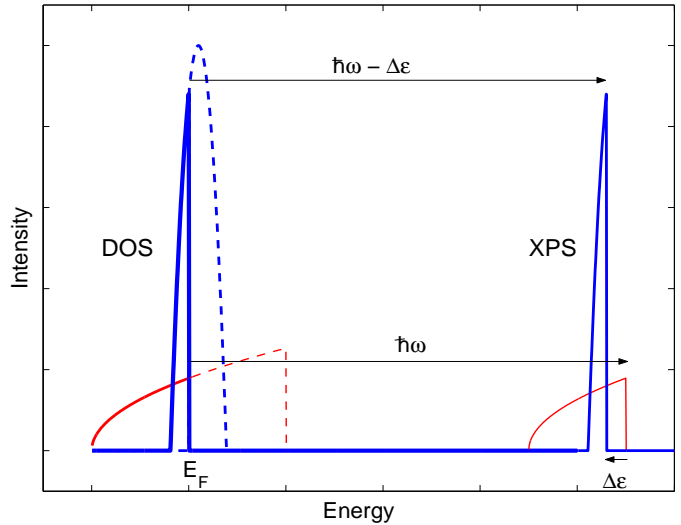}
  \end{center}
  \caption{Schematic picture of the XPS process for a light RE, where the f-band
  is only partially filled. The left hand side of the figure shows the narrow f-band DOS by the
  heavy line (blue) and the wide low-DOS of the itinerant sd-band by the thin line (red) up to $E_F$. The unoccupied
  DOS of these bands are shown by the broken lines. The XPS image would look as in the right hand
  side. The photon energy $\hbar\omega$ is assumed to excite the sd-band with no relaxation, and a clear Fermi
  break. In the excitation process for the f-electron there is a shift ($\Delta\epsilon$) because 
  of the relaxation, so the image of the f-bands appears below the Fermi break of the itinerant band. The
  experimental value of Hubbard $U$ is in ref. \cite{lang} interpreted to be equal to $\Delta\epsilon$.
  In BIS the values of $\Delta\epsilon$ are generally of opposite sign, and the
  (unoccupied) f-band appears to be above the Fermi break.} 
  \label{fig5}
  \end{figure}

\begin{figure}[h]
  \begin{center}
  \includegraphics[width=9cm,height=7cm]{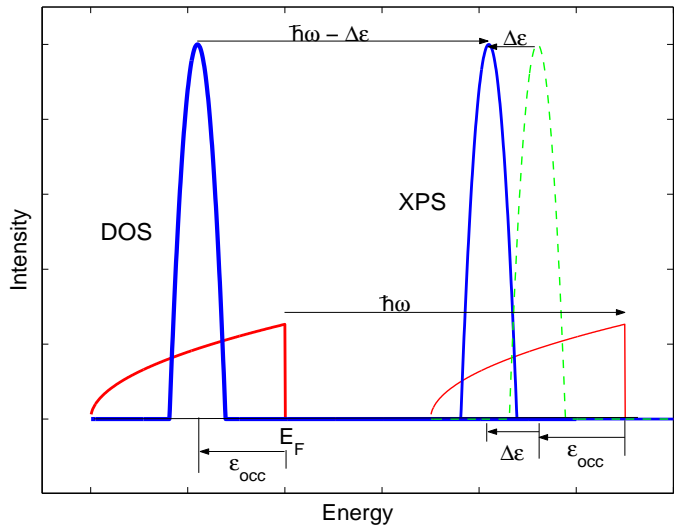}
  \end{center}
  \caption{Schematic picture of the XPS process for RE where the f-band is completely
  filled at an energy $\epsilon_{occ}$ below $E_F$.
  The left hand side shows the occupied DOS of the f- and sd-bands with notations as in Fig. \ref{fig5}. 
  The XPS image, at the right hand side, has the f-band peak down-shifted relative to the sd-band
  by $\Delta\epsilon$. The (green) broken line would be the f-band image without relaxation,
  $\epsilon_{occ}=0$, as from Koopmans theorem. From
  experimental observations of the f-bands \cite{lang} it is tempting, but incorrect, to associate a Hubbard $U$ with
  $\Delta\epsilon + \epsilon_{occ}$, since only $\Delta\epsilon$ is due to many-body
  electron-hole interaction. } 
  \label{fig6}
  \end{figure}

A large part of the energy renormalization is seen directly
in the kinetic term. The remaining non-excited f-electrons on the atom involved in the XPS process
are typically moving closer to $E_F$ and thereby modifying the total energy. 
The energy of the excited electron will mostly be modified downwards
compared to what the Koopmans result would give (see Figures \ref{fig5} and \ref{fig6}).
There is a reduction of the Coulomb and exchange-correlation energies because of the 
screening of the hole. From the values of $\Delta\epsilon$ it seems as if the f-electrons were more
bound, being deeper in energy, compared to what would be expected from the
ground state DOS and the Koopmans theorem. Exceptions to this
are Eu, Yb and Lu, where the f-bands (majority or both) are filled.
The minority f-band well above $E_F$ in Eu is empty, and cannot do screening of the induced hole in the majority band.
In Yb and Lu both f-bands are filled, and screening is also limited.
Hence, the f-band of the remaining electrons move more easily in energy, and in the end 
 it seems as if the f-band move upwards in the three materials. 
The renormalization appears too large in Eu, 
or the LSDA puts its ground state f-level too high in energy. The feasibility to describe
localized f-bands by LSDA have not been much tested because of the discrepancies between bands, spectroscopy
and presumed correlation.  
 
The situation in Gd might be expected to be similar, but its
minority f-band is somewhat occupied, and screening is possible where a fraction of a minority f-electron
replaces the hole in the majority band. The trend 
for Gd and the other RE elements are the same. The measured XPS intensities in Gd (and Eu) \cite{lang} are quite narrow
(of the order eV) due to the absence of large multiplet/SO-splittings, and they are easier to compare with
the calculations. The addition of the band energy $\epsilon_{occ}$ in Table \ref{table1}, -4.2 eV, and the relaxation
energy $\Delta \epsilon^{XPS} = -1.6$ eV in Table \ref{table2}, puts the observed f-band at about -6 eV, which is
in better agreement with the measured peak at -8 eV \cite{lang} than the ground state energy. 
In Eu the correction
of relaxation is positive and puts the peak even above $E_F$, while the only positive corrections for other
elements, in Yb and Lu, makes the agreement with experiment better (bands at -9.3 and -13.5 are corrected upwards
to about -7 and -10 eV, while experimentally they are found as SO-splitted peaks at -2 and -8 eV, respectively).
The measured spectra for the other elements are wider. Nevertheless, there are clearly improved comparisons between
the relaxed band positions and the band centers called $\Delta_-$ extracted from experiments \cite{lang},
for Dy through Tm (-5.3, -4.2, -4 and -5 eV, compared to -3.2, -2.4, -1.7 and
-0.8 eV without relaxation correction, and -3.9, -4.9, -4.7 and -4.6 eV
from experiment \cite{lang}). The corrections for the light elements (Pr, Nd, Sm) are not large, which
suggests that a break in intensity should be found close to $E_F$. Such breaks are seen in the the
XPS intensities, even if the main multiplet peaks are found at lower energy \cite{lang}.

The total energy changes in the BIS process are in general smaller than for XPS. 
There are upward renormalizations of the empty f-bands, which however are quite small
in comparison to the energy of the f-band itself. The observed peaks near 1 and 4 eV in Sm \cite{lang},
agree well with the corrected majority and minority band centers, while for Pr and Nd the energy renormalization is 
underestimated. Gd has the largest correction, and suggests a peak at 3 eV above $E_F$ (instead of about 0.5 eV
for the band), compared to the
BIS-observation at 4 eV \cite{lang}. The observed positions for the elements Tb through Tm, summarized
by the $\Delta_+$-parameters in ref. \cite{lang}, go roughly from 2.8 for Tb to 1.1 eV for Tm. The calculations
show the same relative trend, but with smaller amplitude, from 1 to 0.2 eV.

Inspections of the measured XPS and BIS intensities show more or less sharp Fermi surface
breaks for all elements, even if many high-intensity peaks are not close to $E_F$ \cite{lang}. This
is a hint that some f-electrons are near $E_F$, and that the experimental information is hardly
contained in single $U$-parameters. Nevertheless,
Lang {\it et al} \cite{lang} listed what they call correlation energies ($U$) as being the difference between the peak positions
in XPS ($\Delta_-$) and BIS ($\Delta_+$). However, quite comparable $U$-values can be seen from
the uncorrelated LDA bands shown in Figs. \ref{figdos1}-\ref{figdos3}. In fact, the peak-to-peak
energies of the ground state bands correspond to the exchange splitting of the f-bands,
since the energies generally come from differences between majority and minority bands. Therefore, it is not correct to
assign the difference in peak positions as coming from correlation, at least not beyond what
is already included in LDA. A better agreement with the observed $U$ is obtained when
the XPS and BIS relaxation energies of Table \ref{table2} are added to the peak energies of the ground state
calculations, see Fig \ref{fig4}, even though the comparison is hampered by the absence of SO and multiplet configurations.

Calculations of on-site correlation have been done by forcing an additional electron to
(or removed from) a f-level, in so-called constrained density-functional calculations \cite{dede,mcma,hybe,schn}. 
The total energy differ typically by 5-10 eV or even more from that of the ground state
in such non-equilibrium calculations, and this energy difference is often used as an $U$-value of correlation. It
is tempting to take the peak-to-peak positions as an experimental value of $U$, since they are of the same order
as the constrained DF values. However, as was discussed above, the origin of the peak-to-peak difference
has very little to do with strong on-site correlation. As indicated schematically in Fig. \ref{fig6}
the f-band is already below $E_F$ (by $\epsilon_{occ}$) in many RE metals. Moreover, DFT includes correlation for
the electron gas, where it is relatively more important at low densities. 
On-site correlation can also be
questioned from other points of view \cite{albe}.
On the other hand, the constrained DF-calculations
of $U$ are technically made in a somewhat similar way as in the present work; An electron is 
forced to go into a non-equilibrium level, and differences in total energy are the key parameters.
But, the present method is tailored to the spectroscopic method, and screening reduces the total
energy differences to what is shown in Table \ref{table2} for $\Delta \epsilon^{XPS}$ and 
$\Delta \epsilon^{BIS}$. As seen, the values are usually 1-2 eV and never 
larger than 5 eV.

\begin{table}[ht]
\caption{\label{table2}
 Calculated relaxation, $\Delta \epsilon$, of f states in (majority) XPS and (minority) BIS. Energies in $eV$.
 Calculated and experimental values of the 
 DOS at $E_F$ in units of states per $eV \cdot atom$, 
 obtained from the one-particle band DOS ($N_{band}$, unbroadened DOS with 1 mRy energy resolution), and
  through the calculated
 total free energy as function of $T$ ($N_{calc}$). Experimental values, $N_{exp}$, are
 obtained from the measured values
of the electronic heat capacity coefficients, $\gamma$, 
in ref. \cite{morr}, except for Gd, which is
 from ref. \cite{hill}.}
  \vskip 2mm
  \begin{center}

  \begin{tabular}{l c c c c c}
  \hline

RE (str.) & $\Delta \epsilon^{XPS}$ & $\Delta \epsilon^{BIS}$ & $N_{band}$ & $N_{calc}$ & $N_{exp}$ \\

  \hline \hline

Pr (hcp) &   -0.5 &  0.4 & 12 & 11 & 11 \\
Nd (hcp) &   -0.3 &  0.3 & 16 & 13 & 24 \\
Sm (hcp) &   -0.1 &  0.1 & 14 & 9.5  & 5.2 \\
Eu (bcc) &    1.4 &  0.3 & 1.2 & $\sim$1 & 5.1  \\
Gd (hcp) &   -1.6 &  1.5 & 2.5 & 2.8 & 1.9 \\
Tb (hcp) &   -2.9 &  0.7 & 7.1 & 12 & 4.4 \\
Dy (hcp) &   -2.1 &  0.5 & 16 & 18  & 7.6 \\
Ho (hcp) &   -1.8 &  0.3 & 28 & 19 & 21 \\
Er (hcp) &   -2.3 &  0.2 & 34 & 18 & -  \\
Tm (hcp) &   -4.2 &  0.1 & 20 & 10 & 9.5 \\
Yb (hcp) &    1.7 & -  & 1.1 & $\sim$1 & 1.2 \\
Lu (hcp) &    3.2 & -  & 0.8 & $\sim$1 & 4.2 \\

  \hline
  \end{tabular}
  \end{center}
  \end{table}
  
\subsection{Electronic specific heat}

The total free energy at elevated $T$, $F_T(T)$, needs in principle also excited state corrections.
However, the state at a moderate $T$ is very close to the true
ground state at $T=0$, because the excitations given by the Fermi-Dirac distribution
are on a very small energy scale. The $\ell$-character of levels being occupied just above $E_F$
is almost identical as in the levels of the removed electron just below $E_F$.
(This is very different from spectroscopy, where high-energy dipole transitions are made between initial
and final states.) 
In the spirit of no excited state corrections, we will compare calculated and measured heat capacities
in order to search for evidence of f-electrons at $E_F$.

The electronic free energy $F_T$ is at low $T$ essentially a quadratic function of $T$,
and the heat capacity, $C_{el} = dF_T/dT$ varies linearly with $T$ and can be extrapolated down to
$T \rightarrow 0$ to get the coefficient $\gamma = C_{el}/T$. The relation to the DOS
is given by \cite{kitt}
\begin{equation}
\gamma = \frac{1}{3} \pi^2 N(E_F) k_B^2 (1 + \lambda)
\end{equation}

The electron-phonon coupling $\lambda$ or other many-body interactions such as spin fluctuations,
can enhance the heat capacity, although usually not drastically. The electron-phonon coupling
is not calculated here. Lattice disorder, due to phonons and zero-point motion (ZPM) of the atoms
in the lattice, has an effect
of smearing of the DOS \cite{fesi,fege,dela,gonz,bron}. The cause is mainly coming from the Madelung term of the
potential. This part of the potential is identical for all unit cells in a perfectly ordered lattice, but
the symmetry is broken in the disordered lattice so that different sites have slightly
different potential, which also vary in time. The potential is a classical quantity.
Electronic states, obtained from quantum mechanics, depend on the classical potential
and hence they depend on disorder \cite{fesi}.  This effect is often neglected although it can largely modify
$N(E_F)$-dependent properties. Here, for very narrow f-states at $E_F$, the result would be a smearing
of fine details of $N(E)$ already at small $T$ due to ZPM. A proper calculation of the quantitative
smearing due to disorder is complicated and is out of the scope of this work. However, we will extract
the electronic specific heat coefficients from the calculated variation of $F_T(T)$. 
These calculations are made for smaller (2 atom) cells. The results confirm that $F_T(T)$ is close to a
quadratic dependence of $T$; $F_T(T) \sim T^{\alpha}$, where $\alpha = 2 \pm 0.3$ depending on the
material. Deviations from the parabolic behavior are coming from the sharp
variations of $N(E)$ near $E_F$, and from $T$-variations of charges and spin.

The DOS near $E_F$ varies rapidly with energy when the f-bands are at the Fermi level.
By using $k_BT \sim$ 2 mRy in the Fermi-Dirac function 
it is possible to simulate a DOS-smearing 
as for ZPM at low $T$ \cite{comm}. Thus, $N_{calc}(E_F) = 6 \Delta F_T / (\pi k_B T)^2$ where the difference
in total energy $\Delta F_T = F_T(T) - F_T(0)$ is calculated selfconsistently in temperature intervals up to $k_B T \sim
2 mRy$. This procedure smears
out noise in the $N(E)$ average around $E_F$, and $\gamma$ includes contributions from possible changes in
charge and spin as function of $T$.
The $N_{calc}(E_F)$ averages are sometimes 
different from the band DOS itself, $N_{band}$ (cf. Table II), because of small peaks/dips
that are smeared out by disorder and imperfections in real lattices. Thus $N_{calc}(E_F)$ is
probably more reliable than $N_{band}$ in Table II. The only cases where $N_{band}$ is
better are for
materials with low $\gamma$, because then the DOS has no peaks/dips at  $E_F$, and the
calculation of $F_T(T)$ and $N_{calc}(E_F)$ is less precise.

The results and comparison with experiment are shown in Table II.
The calculated values are in general comparable with the observed values \cite{morr,hill,loun}.
There is no general trend that the f-bands in the ground states should be far away
from $E_F$, since all $\gamma's$ then would be of the same order as for Eu, Yb or Lu. The calculated
f-bands are mostly too narrow, since no SO or multiplet
structures are taken into account.
This explains why the calculated DOS and $\gamma$'s are generally somewhat large in
comparison with experiment. But it is interesting to note that none of the calculated
values is by far too large compared to observation, which would have been the case if the band calculation incorrectly had
put f-electron states at $E_F$. For instance, Table II shows that without f-electrons at $E_F$ one expects
that $N_{calc} \sim$ 1 $(eV \cdot atom)^{-1}$, but when $N_{exp}$ are 10-15 times larger
one can assume that the f-band is at $E_F$ for such RE.
Only Eu, Gd, Yb and Lu have both spin f-bands away from $E_F$ in
the DFT ground states, and their measured $\gamma$'s are also smallest among these RE elements. The highest
$\gamma$'s are measured for Nd and Ho, which also have large calculated $N(E_F)$. 
Eu is unique with a calculated $N(E_F)$ significantly smaller than from experiment. The reason could be
that SO-coupling in combination with a majority f-band rather close to $E_F$ brings more states
to the Fermi level.
In general there is a good correlation between measured $\gamma$'s
and calculated $N(E_F)$ even though enhancing effects of $\lambda$ are neglected. 
Such enhancements should improve the comparison with experiment in the RE without f-electrons
at $E_F$ (Gd, Yb and Lu). It is not clear why large $\lambda$'s seem not to be needed
for the other RE metals with high DOS at $E_F$.
The large $\gamma's$
for most RE elements are compatible with f-electrons (without large enhancements) at the Fermi level of the ground state.

\section{Conclusion}

Observed energy differences in peak-to-peak positions in XPS and BIS
spectra are not measuring on-site correlation $U$, because the DFT ground state positions of 
the f-bands depend more on exchange splitting and conventional 
potential terms. A reasonable comparison can already be made between observed XPS and BIS intensities and
f-band energies of the DFT ground state. For instance, the majority f-bands in Pr, Nd and Sm cross $E_F$,
and discontinuities are seen at $E_F$ in the spectra. Relaxation effects, calculated for the proper
mechanisms of the spectroscopic method, will in general improve the comparison with
experiment by lowering the energies of the XPS peaks and move BIS peaks to higher energy. The effect
is strongest for the bands that do not cross $E_F$, and improves considerably the comparison
between theoretial and observed band centers, at least from what can
be concluded from the band results without SO coupling and multiplets. 

Further improvements of the method,
like representing the high energy state by a band state instead
of the completely delocalized free-electron state, would normally improve XPS results, since the
total energy should be able to relax to a lower value. It has not been tested if potential corrections
based on the generalized gradient
approximation \cite{gga} can lead to improvements for the excited states. 
Ground state properties are usually improved by using GGA, as least for transition metals \cite{bmj}.

Electronic specific heat data compare reasonably well with the DFT results for the ground states,
i.e. where large contributions come from high f-electron DOS at the Fermi level. This implies, together
with the spectroscopic data, that unfilled f-electron bands cross $E_F$, but that they
may appear broadened and shifted away from $E_F$ by the experimental probe. Even if f-electrons
have a large DOS at $E_F$, as in DFT bands, it is not clear that they should be determining for the
electric resistivity, because of their low Fermi velocity. Scattering mechanisms also make
this problem complex. 
At this point we conclude that f-electron energies are easily renormalized in the spectroscopic
process, and that standard LSDA band structures compare reasonably well with electronic
specific heat data at low $T$ of RE elements.
Additional strong on-site correlation energy shifts, of the order 5-10 eV, 
of the f-bands would destroy the agreement. Detailed comparisons between bands and spectroscopies
require energy renormalizations and matrix elements in addition to SO-coupling and multiplets.
Other corrections to DFT potentials are needed for an understanding of metal-insulator
transitions and anti-ferromagnetism, like in the undoped cuprates \cite{tj4}. Solutions
to such problems are not proposed here.

I acknowledge useful discussions with B. Barbiellini.


\begin{thebibliography}{99}
\bibitem{note1} The energy of the final state is not defined precisely, it just has
to be high above $E_F$, at $E_F + \hbar \omega$.
\end{thebibliography}
  \end{document}